%Paper: gr-qc/9508062
%From: <CAVAGLIA@to.infn.it> ("Marco Cavaglia' INFN Sez. di Torino")
%Date: Wed, 30 AUG 95 15:10:27 +0000

%%%%%%%%%%%%%%%%%%%%%%%%%%%%%%%%%%%%%%%%%%%%%%%%%%%%%%%%%%%%%%%%%%%%%%%%%
%%                          Tex instructions                           %%
%%%%%%%%%%%%%%%%%%%%%%%%%%%%%%%%%%%%%%%%%%%%%%%%%%%%%%%%%%%%%%%%%%%%%%%%%

\font\titolino=cmbx10
\font\tsnorm=cmr10

\font\tscorsp=cmti9
\magnification=1200

\hsize=148truemm
%\hsize=138truemm
\hoffset=10truemm
\parskip 3truemm plus 1truemm minus 1truemm
\parindent 8truemm
\newcount\notenumber

\def\PRD#1#2#3{Phys.\ Rev.\ D #1 (#2) #3}
\def\PRL#1#2#3{Phys.\ Rev.\ Lett.\ #1 (#2) #3}
\def\NPB#1#2#3{Nucl.\ Phys.\ B #1 (#2) #3}

\def\IJMPA#1#2#3{Int.\ J.\ Mod.\ Phys.\ A #1 (#2) #3}

\def\MPLA#1#2#3{Mod.\ Phys.\ Lett.\ A #1 (#2) #3}

\def\E{Ein\-stein}
\def\Sc{Sch\-warz\-sch\-ild}

\def\bh{bla\-ck \-ho\-le}

\def\d{\partial}
\def\ra{\rightarrow}
\def\h{\hat}
\def\de{\Delta_{FP}}
\def\pa{p_a}
\def\pal{p_{\alpha}}
\def\dpa{\partial_{p_a}}

\def\ks{Kan\-tow\-ski\--Sa\-chs}

\def\note{\advance\notenumber by 1 \footnote{$^{\the\notenumber}$}}
\def\ref#1{\medskip\everypar={\hangindent 2\parindent}#1}
\def\beginref{\begingroup
\bigskip
\leftline{\titolino References.}
\nobreak\noindent}
\def\endref{\par\endgroup}
\def\beginsection #1. #2.
{\bigskip
\leftline{\titolino #1. #2.}
\nobreak\noindent}
\def\beginappendix #1.
{\bigskip
\leftline{\titolino Appendix #1.}
\nobreak\noindent}
\def\beginack
{\bigskip
\leftline{\titolino Acknowledgments}
\nobreak\noindent}

%\nopagenumbers
\def\frac#1#2{{{#1}/over{#2}}}
%%%%%%%%%%%%%%%%%%%%%%%%%%%%%%%%%%%%%%%%%%%%%%%%%%%%%%%%%%%%%%%%%%%%%%%%%
%%                       End of instructions                           %%
%%%%%%%%%%%%%%%%%%%%%%%%%%%%%%%%%%%%%%%%%%%%%%%%%%%%%%%%%%%%%%%%%%%%%%%%%
%%%%%%%%%%%%%%%%%%%%%%%%%%%%%%%%%%%%%%%%%%%%%%%%%%%%%%%%%%%%%%%%%%%%%%%%%
%%                             Title                                   %%
%%%%%%%%%%%%%%%%%%%%%%%%%%%%%%%%%%%%%%%%%%%%%%%%%%%%%%%%%%%%%%%%%%%%%%%%%
\null
\vskip 5truemm
\rightline {DFTT 50/95}
\rightline{August 10, 1995}
\vskip 15truemm
\centerline{\titolino QUANTIZATION OF THE SCHWARZSCHILD BLACK HOLE}
\vskip 10truemm
\centerline{\tsnorm Marco Cavagli\`a$^{(a,d)}$,
Vittorio de Alfaro$^{(b,d)}$ and Alexandre T. Filippov$^{(c)}$}
\bigskip
\centerline{$^{(a)}$\tscorsp SISSA - International School for Advanced
Studies,}
\smallskip
\centerline{\tscorsp Via Beirut 2-4, I-34013 Trieste, Italy.}
\bigskip
\centerline{$^{(b)}$\tscorsp Dipartimento di Fisica
Teorica dell'Universit\`a di Torino,}
\smallskip
\centerline{\tscorsp Via Giuria 1, I-10125 Torino, Italy.}
\bigskip
\centerline{$^{(c)}$\tscorsp Joint Institute for Nuclear Research}
\smallskip
\centerline{\tscorsp R-141980 Dubna, Moscow Region, Russia.}
\bigskip
\centerline{$^{(d)}$\tscorsp INFN, Sezione di Torino, Italy.}
\vskip 15truemm
\centerline{\tsnorm ABSTRACT}
\begingroup\tsnorm\noindent
%%%%%%%%%%%%%%%%%%%%%%%%%%%%%%%%%%%%%%%%%%%%%%%%%%%%%%%%%%%%%%%%%%%%%%%%%
%%                            abstract                                 %%
%%%%%%%%%%%%%%%%%%%%%%%%%%%%%%%%%%%%%%%%%%%%%%%%%%%%%%%%%%%%%%%%%%%%%%%%%
We quantize by the Dirac -- Wheeler--DeWitt method the canonical
formulation of the \Sc\ black hole developed in a previous paper. We
investigate the properties of the operators that generate rigid
symmetries of the Hamiltonian, establish the form of the invariant
measure under the rigid transformations, and determine the gauge fixed
Hilbert space of states. We also prove that the reduced quantization
method leads to the same Hilbert space for a suitable gauge fixing.
%%%%%%%%%%%%%%%%%%%%%%%%%%%%%%%%%%%%%%%%%%%%%%%%%%%%%%%%%%%%%%%%%%%%%%%%%
%%                             Address                                 %%
%%%%%%%%%%%%%%%%%%%%%%%%%%%%%%%%%%%%%%%%%%%%%%%%%%%%%%%%%%%%%%%%%%%%%%%%%
\vfill
\leftline{\tsnorm PACS: 04.20.-m, 04.60.Ds, 04.70.-s.\hfill}
\smallskip
\hrule
\noindent
\leftline{E-Mail: CAVAGLIA@TO.INFN.IT\hfill}
\leftline{E-Mail: VDA@TO.INFN.IT\hfill}
\leftline{E-Mail: FILIPPOV@THSUN1.JINR.DUBNA.SU\hfill}
\endgroup
\vfill
\eject
\footline{\hfill\folio\hfill}
\pageno=1
%%%%%%%%%%%%%%%%%%%%%%%%%%%%%%%%%%%%%%%%%%%%%%%%%%%%%%%%%%%%%%%%%%%%%%%%%
%%                            text                                     %%
%%%%%%%%%%%%%%%%%%%%%%%%%%%%%%%%%%%%%%%%%%%%%%%%%%%%%%%%%%%%%%%%%%%%%%%%%
\beginsection 1. Introduction.
Recently a good deal of work has been dedicated to the canonical formulation of
the
spherically symmetric gravity and to its quantization [1-3].
In Ref.\ [3] (hereafter referred as I) we developed a canonical approach
to the study of the spherically symmetric metric by proposing a
foliation in the radial parameter $r$ and considering the Lagrangian
coordinates as functions only of $r$. Thus this leads to a structure of
minisuperspace.  The theory is of course endowed with a gauge invariance
(reparametrization in $r$) and a constraint. In I we have developed the
theory and  shown a number of  points that we recall briefly.

We expressed the \E\ equations as a canonical system in a finite, $2\times
2$ dimensional phase space.  The gauge transformation is integrable; in
particular the solution of the equations of motion is the \Sc\ solution.
This property allows the identification of the canonical quantity that
corresponds to the \Sc\ mass. There is an interesting algebraic structure
of three gauge invariant canonical quantities, whose physical meaning was
clarified, that form an affine algebra.

We also started to investigate the Dirac -- Wheeler--De Witt (WDW)
quantization discussing the
general form of the solutions and showed that they are oscillating in the
classically allowed regions and exponentially decreasing in the forbidden
regions.  We briefly discussed the form of the eigenfunctions of the mass
operator and of the  generator of dilatations.  We noted that a set of
solutions coincides with that of \ks\ (KS) wormholes [4,5]. This is hardly
surprising. The geometry inside the horizon of a black hole coincides
with the KS geometry, and further the foliation parameter $r$ is
timelike inside
the horizon, so what we expose here is, for the part internal to the
horizon, isomorphic to the theory of the KS spacetime. This property
enforces the much discussed possibility that a black hole can be
connected to a KS wormhole [6].

What remained to be done was the determination of the measure in the
inner product and the gauge fixing with the consequent establishment of
a positive definite Hilbert space. This is essentially the content of
the present paper.

We start by introducing the classical Lagrangian and Hamiltonian and
integrate the gauge transformations and rigid symmetries. Then we carry
on the construction of the quantum theory. We start with the Dirac
method and establish the WDW equation.

The request of preserving at the quantum level both the gauge invariance
and the classical rigid symmetries, together with the
support properties of the variables used as quantum coordinates,
determines completely the quantum measure and fixes the representation
of the quantum operators.  We identify the solutions of the WDW equation
that are eigenfunctions of
the operators corresponding to the most important invariants of the
classical theory.
A Fourier transform gives the solutions in the configuration space already
found in I using the covariant measure.

Up to this point there has been no gauge fixing
nor definition of a norm. We then fix the gauge by
defining the inner product by the Faddeev--Popov (FP) procedure [7] and
prove the existence of a class of gauges.  This leads to a positive
definite Hilbert space.

We may also start by first fixing the gauge in the classical frame by a
suitable canonical gauge fixing identity that contains the coordinate
$r$ (for the method, see e.g.\ [8]); the results coincide with those
obtained by the Dirac method.

Having a positive definite Hilbert space, we are able to prove that, due
to the support properties of its conjugate variable, the hermitian
operator corresponding to the \Sc\ mass in the gauge fixed, positive
norm, Hilbert space is not self--adjoint,
while its square is a self--adjoint
operator with positive eigenvalues, analogously to what happens for the
radial momentum in ordinary quantum mechanics.

Of course in the classical theory the mass is perfectly defined. Again
take the example of the radial momentum: although it is not a
self--adjoint operator in the Hilbert space, a classical radial momentum
is defined, namely $p_r=m \dot r$, and its square is self--adjoint. This
difference between classical and quantum behavior is due to the fact
that a classical canonical quantity is a purely local entity while the
definition of a self--adjoint operator conveys general informations
about the Hilbert space.
Alternatively, one may think that the operator corresponding to the mass
of the black hole should be defined in a different way.

No quantization of the mass is required by the quantum
theory as it stands. Quantization of the mass can be surely achieved by
a modification of the boundary conditions and/or of the original
Hamiltonian, and we point in the conclusions how it could be carried on
in a gauge invariant way. It would be worth exploring this weird and
fascinating possibility.  It would also be interesting to introduce
matter degrees of freedom with the aim of summing  over degrees of
freedom in order to compute entropy
(hopefully, this may allow us to define a more satisfactory operator
for the mass of the black hole).
\beginsection 2. Classical Theory.
In this section we summarize the main classical results obtained in I
with some minor changes in notation and some
added considerations about the symmetries of the action.

The spherically symmetric line element can be written in
the following convenient form
$$ds^2=-4a(r)dt^2+4n(r)dr^2+b^2(r)d\Omega^2_2 \,,\eqno(2.1)$$
where $a$, $n$, and $b$ only depend on the radial coordinate $r$ and
$d\Omega^2_2$ is the line  element of the two-sphere.

We allow in principle for changes of signs in the metric tensor.
Depending on the sign of $a$ and $n$, the coordinates $t$ and $r$ may be
timelike and spacelike or vice versa. $n(r)$ plays essentially the role of the
$r$-lapse function and it is just a Lagrange multiplier in the action
enforcing the constraint that generates reparametrizations of $r$.

We start from the action
$$S={1\over 16\pi G}\int_{V_4} d^4x \sqrt{-g}~(R+2\Lambda)~-~
{1\over 8\pi G}\int_{\d V_4} d^3x~\sqrt{h}~{\bf K}\,.\eqno(2.2)$$
Introducing the Ansatz (2.1) one obtains
$$S=\int_{t_1}^{t_2} dt\int_{r_1}^{r_2} dr~{\cal L}(a,b,l)\,,\eqno(2.3)$$
where the Lagrangian ${\cal L}$ is
$${\cal L}=2l\left({\dot a b\dot b\over l^2}+{a\dot b^2\over l^2}+
{1+\Lambda b^2\over 4}\right) \,.\eqno(2.4)$$
In Eq.\ (2.4), dots denote differentiation with respect to $r$ and we have set
$G=1$. The Lagrangian multiplier $l$ is given by
$$l(r)=4\sqrt{an}\,.\eqno(2.5)$$
Note that in I the Lagrangian multiplier was chosen as $l(r)=\sqrt{an}/2b^2$.
Also (2.1) and (2.4) are different. We have preferred the present definitions
as they lead to some simplification in the Hamiltonian treatment.
All the results of the present paper, both classical and quantum, are of
course identical.

As already pointed in I, the Lagrangian must be real, so $a$ and $n$
have the same sign and thus the line element (2.1) has lorentzian
signature everywhere: as we will see in a moment, on the classical
solutions positive values of $a$ represent the exterior of the \bh\ and
negative values of $a$ represent the region inside the horizon.

The Hamiltonian $\cal H$ can be calculated in the usual way by
defining canonical $r$--momenta as
$$\eqalignno{&p_a={2b\dot b\over l}\,,&\hbox{(2.6a)}\cr\cr
&p_b={2\over l}(\dot ab+2a\dot b)\,.&\hbox{(2.6b)}\cr}$$
We have:
$$\eqalignno{&{\cal H}=lH\,,&\hbox{(2.7a)}\cr
&H={1\over 2b^2}[p_a(bp_b-ap_a)]-{1\over 2}(1+\Lambda
b^2)\,,&\hbox{(2.7b)}\cr}$$
where $H$ is the generator of $r$-reparametrizations (gauge transformations)
that we will simply call the ``Hamiltonian'' of the system.
As a consequence of the form of $\cal H$ we have the constraint
$$H=0\,,\eqno(2.8)$$
which expresses the invariance under $r$--reparametrization.

Let us set from now on $\Lambda=0$; the case of non zero cosmological
constant will be examined in Appendix B. The Hamiltonian (2.7b) has very
interesting invariance transformations; first, the gauge transformations
generated by $H$ (denoted as ${\cal H}_h$)
$$\eqalignno{&\delta q_i =\epsilon {{\d H} \over {\d p_i}} = \epsilon\bigl[
q_i,H \bigr]_P\,,&\hbox{(2.9a)}\cr\cr
&\delta p_i =-\epsilon {{\d H} \over {\d q_i}} = \epsilon\bigl[
p_i,H \bigr]_P\,,&\hbox{(2.9b)}\cr\cr
&\delta l = {{d \epsilon} \over {dr}}\,,&\hbox{(2.9c)}\cr\cr}$$
can be integrated explicitly. For $H=0$ the result is:
$$\eqalignno{&b \ra \bar b=b+h(r)~{{\pa}\over{2b}}\,,&\hbox{(2.10a)}\cr
&\pa \ra \bar \pa = \pa + h(r)~ {{\pa^2}\over{2b^2}}\,,&\hbox{(2.10b)}\cr
&a \ra \bar a = a + {{N}\over{b^2}}~ {{h(r)/2}\over{1+h(r)\pa/2b^2}}\,,
&\hbox{(2.10c)}\cr
&p_b \ra \bar p_b = p_b + {{J}\over{b^2}}~ {{h(r)/2}\over{1+h(r)\pa/2b^2}}
\,,&\hbox{(2.10d)}\cr
&l(r) \ra \bar l(r) = l(r) + {{dh}\over{dr}}\,,&\hbox{(2.10e)}\cr}$$
where $J$ and $N$ are gauge invariant quantities defined below. Note the
simplicity of the gauge transformations of $b$ and $\pa$. This fact will
be exploited later.

We have three gauge invariant canonical quantities, namely
$$\eqalignno{&I~=~b/p_a\,,&\hbox{(2.11a)}\cr
&J~=~2b-p_ap_b+4bH\,,&\hbox{(2.11b)}\cr
&N~=~IJ~=~bp_b - 2ap_a\,.&\hbox{(2.11c)}\cr}$$
$I$, $J$, $N$ play a fundamental role in the theory. The algebra of
$H,I,J,N$ is:
$${\vbox{
\offinterlineskip\halign{\strut
\qquad#\qquad&\qquad#\qquad&\qquad#\qquad\cr\cr
$\bigl[I,H\bigr]_P = 0\,,$&$\bigl[J,H\bigr]_P = 0\,,$
&$\bigl[J,I\bigr]_P = 1\,,$\cr\cr
$\bigl[N,H\bigr]_P = 0\,,$&$\bigl[N,I\bigr]_P = I\,,$
&$\bigl[N,J\bigr]_P = -J\,.$\cr\cr}}}\eqno(2.12)$$
We write also the unconstrained solution of the equations of motion; for
$H=0$ they have of course the same content as the gauge equations (2.10).
We have
$$\eqalignno{&b={\tau\over 2I}\,,&\hbox{(2.13a)}\cr
&p_a={{b}\over {I}}\,,&\hbox{(2.13b)}\cr
&a=I^2\left(2H + 1-{{J}\over{b}}\right)\,,&\hbox{(2.13c)}\cr
&p_b=I\left(4H + 2-{J\over b}\right)\,,&\hbox{(2.13d)}\cr
&\tau=\int_{r_0}^r l(r)~dr\,,~~~~~~l(r)>0\,.&\hbox{(2.13e)}\cr}$$
$\tau$ will be chosen positive without loss of generality.

Eq.\ (2.13c) corresponds to the \Sc\ solution
if we set $H=0$. Then $a$ vanishes for
$b=J$ and so $J/2\equiv M$ is the classical canonical expression of the
\Sc\ mass M. Again from (2.13c), remembering (2.1), we see that $T\equiv
2I$ is the ratio between proper and coordinate time in the asymptotic
region $b\to\infty$.

There is a very important point concerning the support of the variables
$b$ and $p_a$. $b$ is positive definite as it is natural since it is
classically a radial variable. Then from the positivity of $\tau$  it
follows from (2.13a) that $I>0$. Also, $p_a$ is positive. These
properties will be essential in the following.

Let us call rigid those symmetries generated by $I$, $J$ and $N$. Any
gauge invariant function of the canonical variables can be written as
$F(H,I,J)$. The requests that it be $N$ and $I$ invariant, or $N$ and
$J$, or $I$ and $J$, are equivalent as they leave us with $F(H)$,
so we need only consider two of the three rigid transformations.

Invariance under rigid transformations will be used in the next chapter
to investigate the quantum measure. For the moment let us write down
these transformations.

The finite transformations
generated by $I$ (denoted as ${\cal I}_f$) are:
$$\eqalign{
&b \ra \bar b = b\,,\cr
&\pa \ra \bar\pa = \pa\,,\cr
&a \ra \bar a = a - f b p_a^{-2}\,,\cr
&p_b \ra \bar b_b= p_b- f\pa^{-1}\,.\cr}\eqno(2.14)$$
The finite transformations generated by $J$ (denoted as ${\cal J}_q$) are:
$$\eqalign{
&b \ra \bar b = {b\over 1-q/I}\,,\cr
&\pa \ra \bar \pa = {\pa\over\left(1-q/I\right)^2}\,,\cr
&a \ra \bar a = a\left(1-q/I\right)^2+{N\over b}q
\left(1-q/I\right)^2\,,\cr
&p_b \ra \bar p_b = p_b\left(1-q/I\right)+{J\over b}q
\left(1-q/I\right)\,.\cr}\eqno(2.15)$$
The finite transformations generated by $N$ (denoted as ${\cal N}_g$)
on the canonical variables are dilatations, due to the form of $N$:
$$\eqalign{
&b \ra \bar b = e^g b\,,\cr
&\pa \ra \bar \pa = e^{2g}\pa\,,\cr
&a \ra \bar a = e^{-2g} a\,,\cr
&p_b \ra \bar p_b = e^{-g} p_b\,.\cr}\eqno(2.16)$$

Now looking at these three sets of transformations and at the gauge
transformations (2.10) we see that the canonical variables $b,p_a$
transform separately under all transformations. These variables will be
most appropriate as coordinates in the quantum case. We also note that
the $J$ transformations may change the sign of $b$, in contrast with our
assumption that $b>0$. Thus we consider as fundamental symmetries
${\cal N}_g$ and ${\cal I}_f$. The analysis of the consequences of relaxing the
condition $b>0$ (unfolding of $b$) will be carried on elsewhere.

Different sets of canonical pairs will be used in what follows.
We may perform a canonical transformation to the new canonical
variables $\{J,I,Y,P_Y\}$, where
$$Y={2b^2\over p_a}\,,~~~~~~~~P_Y=H\,.\eqno(2.17)$$
This choice is motivated by their invariance properties: $I,J,H$ are
gauge invariant and $Y$ behaves in the simplest way under gauge
transformations ${\cal H}_h$.
For completeness we give  the generating function of the canonical
transformation:
$$F~=~ {2ab^2\over Y}-{Y J\over 2b}+{1\over 2}Y\,. \eqno(2.18)$$
We can use alternatively $N=IJ$ and $p_N=\ln I$ instead of $J$ and $I$.
Using the canonical variables $\{J,I,Y,P_Y\}$, the Hamiltonian reads simply
$${\cal H}=lP_Y\,.\eqno(2.19)$$
We list their transformation properties under gauge and rigid
transformations:
\medskip\noindent
$\bullet~{\cal I}_f$:
%I \ra
$$\eqalign{&I \ra \bar I = I\,,\cr
&J \ra \bar J=J+f\,,\cr
&Y \ra \bar Y=Y\,,\cr
&P_Y \ra \bar P_Y=P_Y\,;\cr}\eqno\hbox{(2.20)}$$
$\bullet~{\cal J}_q$:
$$\eqalign{&I \ra \bar I = I-q\,,\cr
&J \ra \bar J=J\,,\cr
&Y \ra \bar Y=Y\,,\cr
&P_Y \ra \bar P_Y=P_Y\,;\cr}\eqno\hbox{(2.21)}$$
$\bullet~{\cal N}_g$:
$$\eqalign{&I \ra \bar I=e^{-g}I\,,\cr
&J \ra \bar J= e^g J\,,\cr
&Y \ra \bar Y = Y\,,\cr
&P_Y \ra \bar P_Y = P_Y\,;}\eqno\hbox{(2.22)}$$
$\bullet~{\cal H}_h$:
$$\eqalign{&I \ra \bar I=I\,,\cr
&J \ra \bar J= J\,,\cr
&Y \ra \bar Y = Y+h(r)\,,\cr
&P_Y \ra \bar P_Y = P_Y\,.\cr}\eqno(2.23)$$
These formulas will be important for the discussion in the next section.

Finally, let us write another set of canonical variables  that will
be used in section 4:
$$\eqalign{&\alpha=\ln|a|\,,\cr
&c=2\sqrt{|a|}b\,,\cr
&p_\alpha=-{N\over 2}\,,\cr
&p_c={p_b\over 2\sqrt{|a|}}\,.\cr}\eqno(2.24)$$

In this case there are two different canonical transformations,
for positive and negative $a$; the Hamiltonian (2.7b) becomes
$$H={1\over 2}\left[\sigma\left(p_c^2-4{p_\alpha^2\over c^2}\right)
-1\right]\,,\eqno(2.25)$$
where $\sigma=a/|a|$.
The gauge transformation laws of these canonical coordinates
are not simple:
$$\eqalign{&\delta p_\alpha=0\,,\cr
&\delta\alpha=-4\epsilon\sigma{p_\alpha\over c^2}\,,\cr
&\delta p_c=-4\epsilon\sigma{p_\alpha^2\over c^3}\,,\cr
&\delta c=\epsilon\sigma p_c\,.\cr}\eqno(2.26)$$
Let us remark that these canonical variables become useless in the case of
non vanishing cosmological constant.
\beginsection 3. Quantization.
There are two approaches to the quantization of gauge systems [7].
The first is the Dirac method that leads in our case to the WDW
equation and needs gauge fixing before being interpreted.
This method has the problem of the choice of the measure and the
related problem of the representation of the operators. This difficulty
is usually overcome by the definition of an invariant measure in
superspace.

The second approach is the canonical gauge fixing method leading
to a classical
reduced phase space where quantization can be carried on as usual and
wave functions have the customary interpretation.

In our treatment of the quantization of the black hole one may
carry on both methods and we will be able to show that they
lead to the same results for correct gauge fixing conditions, thus
proving the equivalence of the two approaches.
Most of this section is dedicated to the Dirac method, devoting the
final part to the discussion of the canonical gauge fixing.

In order to
implement the Dirac procedure, the first main problem we meet is the
choice of the measure in superspace and, as consequence, the choice of
the variables to be used for the wave functions. We start with the
formal commutation relations
$$\eqalignno{&\bigl[a,p_a\bigr] = i\,,&\hbox{(3.1a)}\cr
&\bigl[b,p_b\bigr] = i\,.&\hbox{(3.1b)}\cr}$$
In order to represent them as differential operators we must first
choose a pair of commuting variables  as coordinates and establish the
form of the (non gauge fixed) measure $d\mu$. The measure $d\mu$ can be
determined by the requirement that it be invariant under the symmetry
transformations of $H$, namely rigid and gauge transformations.

We shall see in section 4 that the wave functions obtained with this
measure are connected by a Fourier transform to the solutions of the
WDW equation that uses the covariant measure in the $a,b$ space.

Let us come back to the algebra of $I$, $J$, $N$ and $H$. This is a
powerful inspiration for physical consequences to be found in the
structure of the gauge fixed positive definite Hilbert space. The $I$,
$J$, $N$ algebra is a dilatation algebra, so it is useful to recall some
important points about the self--adjointness of the dilatation operator
[9].

Let us consider a realization of the dilatation algebra on differentiable
functions of a single variable $\xi$. If the support of the eigenvalues
of both $\h\xi$ and $\h p_{\xi}$ extends from $-\infty$ to $\infty$, then
$\h\xi$ and $\h p_{\xi}$ are self--adjoint while the dilatation operator
$\h D=(\xi p_\xi+p_\xi \xi)/2$ is not self--adjoint. If instead
for instance $\xi\geq 0$, then (as typical for radial variables) the
dilatation generator is self--adjoint and the conjugate momentum
$\h p_{\xi}$ is not.
So we expect that the support of the variables in the present problem
will be the key to the properties of the Hilbert space.

In order to determine the quantum measure, we require that the measure
be invariant under rigid and gauge transformations. We choose the
${\cal N}_g$ and ${\cal I}_f$ form of the rigid symmetries, (2.14,16),
because they preserve the
sign of $b$. Then the measure is (we denote by $x,~j,~y$ the continuous
eigenvalues of $\hat I$, $\hat J$, $\hat Y$):
$$d\mu(x,y)~=~{dx\over x}~dy\,.\eqno(3.2)$$
This measure makes sense as we have seen that classically $I>0$ since
both $p_a,b>0$. We cannot use the ${\cal N}_g$ and ${\cal J}_q$ form of
the rigid
symmetries, as they change the sign of $b$ (and of $I$). The choice of
implementing the rigid symmetries ${\cal N}_g$, ${\cal J}_q$ implies that
$b$ becomes
negative, for which there is no basis. In that case the invariant
measure would be
$$d\mu(j,y)~=~{dj\over j}~dy\,,\eqno(3.3)$$
that requires $j>0$.  Of course one
could argue that $j>0$ because we have to exclude negative masses, but
this choice would introduce an external criterion into the discussion.
As we will see in a moment, the measure (3.2) implies that the
operator $\hat J$ is not self--adjoint.

The measure (3.2) can be obtained through different considerations, i.e.\
using as variables the pair $\{b,\pa\}$ whose behaviour is simple under
both rigid and gauge transformations. This pair of non conjugate
variables is a basis for a representation of the gauge group and
therefore $b$ and $\pa$ are good candidates as coordinates in the wave
functions. It is straightforward  to determine the form of the invariant
measure in this representation. Let
$$d\mu(b,\pa)~=~F(b,\pa)~db d\pa\,;~~~~
d\bar \mu(\bar b,\bar\pa)~=~F(\bar b,\bar \pa)~d\bar b
d\bar \pa\,.\eqno(3.4)$$
We have:
$$d\bar \mu~\approx~d\mu \left(1 + \Delta + f_b \delta b +
f_{\pa} \delta \pa\right) db d\pa\,,\eqno(3.5)$$
where
$$f~=~\ln F\,,~~~~ f_k~ =~\d_k f\,,
{}~~~~{{\d(\bar b,\bar \pa)}\over{\d (b,\pa)}}-1
\approx\Delta\,,\eqno(3.6)$$
and
$$\Delta ~ = ~ 3g + {{h\pa}\over{2b^2}}\,.\eqno(3.7)$$
The condition of invariance determines completely $F$:
$$F~=~{{b}\over{\pa^2}}\,.\eqno(3.8)$$
The measure invariant under the continuous transformations
${\cal N}_g$, ${\cal I}_f$ that leave $H$ invariant is thus
$$d\mu(b,\pa)~=~ {{bdb~ d\pa}\over{ \pa^2}}\,.\eqno(3.9)$$
It is immediate to see that it coincides with (3.2).

Let us consider for a moment the set of rigid transformations ${\cal N}_g$
and ${\cal J}_q$. In spite of the simple transformation properties of $b,p_a$
under them, it is easy to see by the above method  that an invariant
measure of the form (3.4) cannot be determined. Furthermore,
the measure (3.3) is invariant under ${\cal J}_q$, ${\cal  N}_g$ and
${\cal H}_h$ but cannot be transformed back to the canonical variables
$\{b,p_a\}$.

So let us go back to
the measure (3.9) or (3.2). We can define the operators $\h H,~\h
N,~ \h J$ both in the $\{b,\pa\}$ and in the $\{x,y\}$ representation.
Using the first pair of coordinates we have the hermitian operators
$$\eqalignno{&\h a~=~ i~ \pa \dpa \pa^{-1}\,,&\hbox{(3.10a)}\cr\cr
&\h p_b~=~-i~ b^{-1/2} \d_b~ b^{1/2}\,.&\hbox{(3.10b)}\cr}$$
Note that $\h H$ is first order in derivatives,
as well as $\h N$ and $\h J$. Using the Weyl ordering we obtain:
$$\eqalignno{&\h H=-i{{p_a}\over{2b^2}} \left(b\d_b+p_a \d_{p_a}\right)
-{1\over 2}\,,&\hbox{(3.11a)}\cr\cr
&\h N=-i\left(b\d_b+2p_a \d_{p_a} \right)\,,&\hbox{(3.11b)}\cr\cr
&\h J=-i{\pa\over b}\left(b\d_b+2\pa\dpa+{1\over 2}\right)\,.
&\hbox{(3.11c)}\cr}$$
Let us first discuss the eigenfunctions of $\h N$. The solution of
$$\h H\Psi=0,~~~~\h N\Psi=\nu\Psi\,,\eqno(3.12)$$
is
$$\Psi_{\nu}(b,p_a)~=~c(\nu)~
b^{-i\nu}~ p_a^{i\nu}~ e^{ib^2/p_a}\,,\eqno\hbox{(3.13a)}$$
or, in terms of $x$, $y$:
$$\Psi_{\nu}(x,y)~=~c(\nu)~x^{-i\nu}e^{iy/2}\,.\eqno\hbox{(3.13b)}$$
The eigenfunctions of the mass operator $\h J$ are the solution of the
equations:
$$\h H\Psi=0,~~~~\h J \Psi=j\Psi\,,\eqno(3.14)$$
namely,
$$\Psi_j(b,\pa)~=~c(j)\sqrt{{b\over\pa}}~e^{ib(b-j)/\pa}\,,
\eqno\hbox{(3.15a)}$$
or, in the $\{x,y\}$ representation:
$$\Psi_j(x,y)~=~c(j)\sqrt{x}~e^{i(y/2-jx)}\,.\eqno\hbox{(3.15b)}$$
For sake of completeness, let us obtain from the differential representation
(3.11) the form of the operators $\h H$, $\h J$, $\h N$ in the $\{x,y\}$
representation:
$$\eqalignno{&\h H~=~P_y~=-i \d_y-{1\over 2}\,,&\hbox{(3.16a)}\cr
&\h J~=~ i \sqrt{x} \d _x {{1}\over{\sqrt{x}}}\,,&\hbox{(3.16b)}\cr
&\h N~=~ i {{\d}\over {\d \ln x}}\,,&\hbox{(3.16c)}\cr}$$
Now in order to progress we have to introduce the gauge fixing via the
FP method [7].
We will prove that there is a class of viable gauges for which there are
no Gribov copies and the FP determinant $\de$ is invariant under gauge
transformations. Indeed, let us suppose that the gauge be enforced by
$$\Phi(x,y)~=~0\,,\eqno(3.17)$$
and let $\Phi$ have the form
$$\Phi(x,y)=\psi(x,y)~\prod_i \bigl(y-\phi_i(x) \bigr)\,,\eqno(3.18)$$
where $\psi\bigl(x,\phi_i(x)\bigr) \not=0$ and $\phi_i(x)\not=\phi_j(x)$
for any $x$. Then
$$\delta(\Phi)= \sum_i \delta\left(y-\phi_i(x)\right) ~\psi_i(x)\,,
\eqno(3.19)$$
where
$$\psi_i(x)= \psi\bigl( x,\phi_i(x)\bigr)
 \prod_{j\not= i} \bigl(\phi_i(x)-\phi_j(x)\bigr)\,.\eqno(3.20)$$
So, finally,
$$\de^{-1}~=~\int dh~\delta\bigl(\Phi(h)\bigr)~=~\sum_i
\bigl(\psi_i(x)\bigr)^{-1}\,.\eqno(3.21)$$
Note that since $x$ is gauge invariant, so is $\de$. The gauge fixed
invariant measure is then
$$\int d\mu(x,y)~ \delta\bigl(\Phi(x,y)\bigr)~\de~=~\int
{{dx}\over{x}} dy~\delta\bigl(\Phi(x,y)\bigr)~\de\,.\eqno(3.22)$$
In our case the most convenient gauge (3.17) is:
$$\Phi(x,y)~=~y - 1~=~{{2b^2}\over{\pa}} - 1\,.\eqno(3.23)$$
This gauge fixing implies obviously $\de=1$ and it determines uniquely
the gauge. Indeed,
$${{2\bar b^2}\over{\bar \pa}} = 1\eqno(3.24)$$
defines uniquely $h=1-2b^2/\pa$.

Now we may discuss the form of the wave functions in the gauge (3.23).
Denote by lower case greek letters the wave functions in the
gauge fixed representation and start from the eigenfunctions of $\h N$.
Choosing $c(\nu)=(2\pi)^{-1/2}$, the gauge fixed
eigenfunctions of $\h N$ are
$$ \psi_{\nu}(x) ~=~{{1}\over{\sqrt{2\pi}}}~x^{-i\nu}\,. \eqno(3.25)$$
They are of course orthonormal in the gauge fixed measure:
$$\left(\psi_{\nu_{2}},\psi_{\nu_{1}}\right)=
\int_0^{\infty}~{{dx}\over{x}}~~ \psi^\star_{\nu_2}(x) \psi_{\nu_1}(x)~=
{}~\delta(\nu_1-\nu_2)\,.\eqno(3.26)$$
Now consider the gauge fixed eigenfunctions of $\h J$:
$$\psi_j(x)~=~c'(j)~\sqrt{x}~e^{-ixj}\,.\eqno(3.27)$$

This makes clear the important point already stressed. It is indeed
immediate to verify that $\h J$ is not self--adjoint in that space.  As
already remarked, the situation is similar to the familiar case of the
radial coordinate $r$ in flat space: its conjugate $p_r$ is not a
self--adjoint operator on the Hilbert space of the Laplace operator,
although it is of course a well defined classical quantity.

If, as it is suggested by the classical correspondence, we
identify $\h J$ with the mass operator, we must conclude that there is
no self--adjoint mass operator in this reduced theory.
In other words, with this definition the mass operator is not an observable.

To conclude this section, let us now investigate the operator $\h J^2$.
In order to be a self--adjoint operator, the eigenfunctions of
$\h J^2$ with eigenvalue $j^2$ must meet one of the two conditions:
$$\lim_{x\ra0}~{\psi_{j^2}^{(1)}(x)\over\sqrt{x}}~=0\,, \eqno\hbox{(3.28a)}$$
or
$$\lim_{x\ra0}~\left[{\psi_{j^2}^{(2)}(x)\over\sqrt{x}}\right]'~=0\,.
\eqno\hbox{(3.28b)}$$
The two separate sets are given of course by ($j>0$)
$$\eqalignno{&\psi_{j^2}^{(1)}(x)~=~ {1\over\sqrt{\pi j}}~\sqrt{x}~
\sin jx\,,&\hbox{(3.29a)}\cr
&\psi_{j^2}^{(2)}(x)~=~ {1\over\sqrt{\pi j}}~\sqrt{x}~ \cos jx\,.
&\hbox{(3.29b)}\cr}$$
Either the set (3.29a) or the set (3.29b) must be chosen. The
eigenfunctions of each set are orthonormal
$$\left(\psi_{j^2_2}^{(k)},\psi_{j^2_1}^{(k)}\right)=\int_0^\infty
{{dx}\over{x}}~\psi_{j^2_2}^{(k)^\star}(x)~\psi_{j^2_1}^{(k)}(x)~
=~\delta(j_2^2-j_1^2)\,,~~~~k=1,2\,.\eqno(3.30)$$
Thus the operator $\h J^2$ is self--adjoint.
The effect of the non self--adjoint operator $\h J$ is to transform the
set (1) into the set (2) and viceversa.

The same results can be obtained by the canonical gauge fixing method
(see [8]) using the gauge fixing condition
$$Y=r\,.\eqno(3.31)$$
This gauge fixing (3.31) corresponds to the ``area gauge'' $b={\rm
const}\,\cdot\,r$ since $Y=2bI$. Indeed, we have $l=1$.
The effective Hamiltonian on the physical shell is
$$H_{\rm eff}=-P_Y\equiv -H=0\,.\eqno(3.32)$$
So the functions do not depend on $r$. Diagonalizing $\h N$ or $\h J$ using
(3.16b,c) one obtains the gauge fixed wave functions (3.25) and
(3.27). This proves the equivalence of the Dirac--WDW and reduced
canonical quantization methods for the gauge fixings that we have
implemented.
\beginsection 4. WDW solutions in the $\bf\{a,b\}$ representation.
We now follow the traditional path of determining the measure by
defining the kinetic part of the Hamiltonian as a Laplace--Beltrami operator.
We use the couple of variables $a,b$. From (2.4) we read the covariant
measure in superspace
$$d\mu(a,b)~=~b~da~db\,.\eqno(4.1)$$
The representation for $\h p_a$ and $\h p_b$ is thus:
$$\hat p_a=-i\partial_a,~~~~~\hat p_b=-i(\partial_b+1/2b)\,.\eqno(4.2)$$
In the $\{\alpha,c\}$ representation the covariant measure is $c~d\alpha dc$
and we have:
$$\hat\pal=-i\partial_\alpha,~~~~~\hat p_c=-i(\partial_c+1/2c)
\,.\eqno(4.3)$$
Using the covariant Laplace--Beltrami ordering for the Hamiltonian (note
that it coincides with the Weyl ordering) the WDW equation becomes:
$$[ab\d_a\d_b-(a\d_a)^2+ab^2]\Psi=0\,,\eqno(4.4)$$
or, in terms of $\alpha$ and $c$:
$$[-(c\d_c)^2+4\d_\alpha^2-\sigma c^2]\Psi=0\,,\eqno(4.5)$$
where $\sigma=a/|a|$. The representations for the operators $\hat J$ and
$\hat N$ are
$$\hat J=\d_a\d_b+2b-{1\over 2b}\d_a\,,\eqno(4.6)$$
$$\hat N=-i(b\d_b-2a\d_a)\,.\eqno(4.7)$$
It is easy to check that, using a different definition of
the Lagrange multiplier, the WDW differential equation (4.4) and the
differential expressions for $\hat J$ and $\hat N$ (4.6,7) remain unchanged.

Now let us discuss the diagonalization of $\hat N$.
We have to discuss separately the cases $a>0$ and $a<0$.
The solutions are:
$$\Psi_{\nu}(a,b)=c(\nu)(-a)^{-i\nu/2}K_{i\nu}(2b\sqrt{-a})
\,,\eqno\hbox{(4.8a)}$$
for $a<0$, where $K_{i\nu}$ is the modified Bessel function of order
$i\nu$ [10] (we have chosen this solution because of its asymptotic
properties for large argument). For $a>0$, we have
$$\Psi_{\nu}(a,b)=c'(\nu)a^{-i\nu /2}C_{i\nu}\bigl(2b\sqrt{a}\bigr)\,,
\eqno\hbox{(4.8b)}$$
where the function $C_{i\nu}$ is any combination of Hankel
functions.
For the $\hat J$ operator, the solutions with eigenvalue $j$ are
$$\Psi_j(a,b)=
{K(j)\over\sqrt{|b-j|}}e^{\pm 2i\sqrt{ab(b-j)}}\,,\eqno\hbox{(4.9a)}$$
in the classically allowed region ($a(b-j)>0$, oscillating behavior), and
$$\Psi_j(a,b)= {K(j)\over\sqrt{|b-j|}}e^{-2\sqrt{ab(j-b)}}\,,
\eqno\hbox{(4.9b)}$$
in the classically forbidden region $a(b-j)<0$, where we have chosen the
decreasing exponential behavior, analogously to (4.8a).

Now we may see that these solutions are the Fourier transforms of the
solutions in the $\{b,p_a\}$ space obtained in the previous section. The
Fourier transform is defined as:
$$\Psi(a,b)=\int_0^\infty {d\pa\over\pa^2}~\Psi(\pa,b)~\pa
e^{ia\pa}\,.\eqno(4.10)$$
Introducing in (4.10) $\Psi_\nu(\pa,b)$ given in (3.13a) and using
Ref.\ [11] (Vol.\ I, p.\ 313, formula (17)), one obtains (4.8a); (4.8b) is
obtained by elementary analytic continuations. Analogously,
introducing (3.15a) one obtains (4.9a) or (4.9b). This proves the
equivalence of the invariant measure (3.9) and of the  representations
(3.10) with the covariant measure (4.1) and representation (4.2).

Possibly there is no gauge fixing in the Dirac method leading to a
positive definite Hilbert space of states $\Psi(a,b)$, analogously to
what happens in the very similar Klein--Gordon (KG) case. Note that if we
use the KG light cone variables $p_{+}$ and $x_{+}/p_{+}$ ($p_{+}=p_0 +
p$, $x_{+} = x_0 + x$, see below Eqs.\ (4.11,12)), we may apply the procedures
of the previous section to the KG case. This fact follows from the canonical
equivalence
of the classical black hole to the classical KG theory. This equivalence
will be discussed in a forthcoming publication.

Let us now discuss the gauge fixing by the canonical method,
i.e.\ by a canonical identity and quantizing in the reduced phase space.
In connection with this it will be interesting to recall a few
interesting facts about the KG theory. Let us consider the relativistic
particle in two dimensions. The Hamiltonian is
$${\cal H}=l(t)H\,,~~~~~H={1\over 2}\bigl(p^2+m^2-p_0^2\bigr)\,.\eqno(4.11)$$
The equations of motion are:
$$\dot x = lp\,,~~~~\dot p = 0\,,~~~~ \dot x_0 = -lp_0\,,~~~~ \dot p_0
=0\,;\eqno\hbox{(4.12a)}$$
$$H = 0\,. \eqno\hbox{(4.12b)}$$
The gauge can be fixed via the canonical method imposing the identity
$$x_0 + t = 0\,. \eqno(4.13)$$
As a consequence, (4.12b) and (4.13) become second class and
the system can
be reduced. Eq.\ (4.13) gives the Lagrangian multiplier $l=1/p_0$. Using
(4.12b) and (4.13) we find
$$H_{\rm eff} = ~ p_0~=\pm\sqrt{p^2 ~+ ~m^2}.\eqno(4.14)$$
In order to have a positive Lagrange multiplier and a sensible quantum
mechanics of a single particle, we have to choose the positive sign in
(4.14). So we end with the reduced space Hamiltonian
$$ H_{\rm eff}~=~\sqrt{p^2 ~+ ~m^2}\,. \eqno(4.15)$$
This is the gauge fixed Hamiltonian of the relativistic particle. The
choice of the $-$ sign in (4.14) would be wrong in quantum mechanics of
a single particle (see e.g.\ [12]). The Schr\"odinger
equation is ($+$ stands for $l>0$)
$$i{\d\over\d t}\psi_+(x,t)~=\hat H_{\rm eff}~\psi_+\,,\eqno(4.16)$$
and the eigenfunctions of $\hat p$ are
$$\psi_+(k;x,t) ~ = ~ (2\pi)^{-1/2} ~ \exp[-it \omega +ikx]\,,
\eqno(4.17)$$
where of course $\omega=\sqrt{k^2+m^2}$.
This is obviously a positive definite Hilbert space since the
Hamiltonian is positive definite and hermitian. Usual quantum mechanics
applies.

Now go back to our problem and discuss the canonical gauge fixing for the
\bh. The discussion parallels that of the KG, as (4.5) is essentially a
KG system in the $\{\alpha,c\}$ representation.

Using the variables $\alpha$ and $c$ and the Hamiltonian
(2.25), the equations of motion are
$$\dot c=l\sigma p_c\,,~~ \dot \alpha=-4l\sigma p_{\alpha}/c^2\,~~
\dot p_{\alpha}=0\,,~~ \dot p_c=-4l\sigma p_{\alpha}^2/c^3\,;
\eqno\hbox{(4.18a)}$$
$$H=0\,.\eqno\hbox{(4.18b)}$$
It is convenient to
choose the gauge fixing canonical identity (analogous to (4.13)):
$$\alpha ~= ~ r \,, \eqno(4.19)$$
(of course with the gauge above $r$ is not the area coordinate).
The effective Hamiltonian is thus:
$$H_{\rm eff} ~ = ~ -\pal ~ = N/2\,, \eqno(4.20)$$
where $\pal$ can be obtained from $H=0$. Hence
$$H_{\rm eff}~ =~ \pm{1\over 2}~ \sqrt{c^2(p_c^2~-~\sigma)}\,.
\eqno(4.21)$$
Note that in the classical motion the argument in the square root never
becomes negative. This is obvious for $a<0$. For $a>0$ it can be seen as
follows: from (2.13c,d) we have the relation $p_b=a/I+I$ and using the
definition of $p_c$ in Eqs.\ (2.24) it follows that
$p_c^2=p_b^2/4a=(a/I+I)^2/4a\geq 1$.

Let us look at the value of the Lagrange multiplier. From (4.19) and
from the equations of motion (4.18) we have
$$ l~=~-{{c^2 \sigma}\over{4p_{\alpha}}}\,. \eqno(4.22)$$
Now, as in the KG case, we impose that $l>0$, that is $\sigma
p_{\alpha}<0$. This means that for $a>0$ we must choose the $+$ sign in
(4.21), while for $a<0$ we have to choose the $-$ sign. Let us
use (4.3) and the covariant ordering. First discuss $a<0$. The
eigenstates of $H_{\rm eff}$ with eigenvalue $E=-\nu/2,~\nu>0$
are obtained by solving the equation
$$[-(c\d_c)^2+~c^2]~\psi_{\nu}(c)~=~{\nu^2}~
\psi_{\nu}(c)\,.\eqno(4.23)$$
The solution is
$$\psi_{\nu}(c)~=~\sqrt{{2\nu\sinh{\pi\nu}}\over\pi^2}
{}~ K_{i\nu} (c)\,.\eqno(4.24)$$
For the case $a>0$ we look for eigenstates of $H_{\rm eff}$
$$[-(c\d_c)^2-~c^2]~\chi_{\nu}(c)~=~{\nu^2}~
\chi_{\nu}(c)\,,\eqno(4.25)$$
with solution ($\nu>0$)
$$\chi_{\nu}(c)~=i\sqrt{\nu\sinh(\pi\nu/2)\over 4\cosh(\pi\nu/2)}
\left[e^{-\pi\nu/2}H^{(1)}_{i\nu}(c)-e^{\pi\nu/2}H^{(2)}_{i\nu}(c)
\right]\,.\eqno(4.26)$$
The above solutions of the Hamiltonian are orthonormal (see Appendix A):
$$\left(\psi_{\nu_1},\psi_{\nu_2}\right)=\int_0^\infty ~{{dc} \over {c}}
{}~\psi_{\nu_1}^\star (c) ~\psi_{\nu_2} (c)~=~ \delta(\nu_1 - \nu_2)\,,
\eqno\hbox{(4.27a)}$$
$$\left(\chi_{\nu_1},\chi_{\nu_2}\right)=\int_0^\infty ~{{dc} \over {c}}
{}~\chi_{\nu_1}^\star (c) ~\chi_{\nu_2} (c)~=~ \delta(\nu_1 - \nu_2)\,.
\eqno\hbox{(4.27b)}$$
These eigenfunctions span positive norm Hilbert spaces.

Now let us solve the Schr\"odinger equation
$$i~ {{\d} \over {\d \alpha}}~ \Psi_+(\nu;c,\alpha)~ = ~H_{\rm eff}~
\Psi_+(\nu;c,\alpha)\eqno(4.28)$$
for the stationary states. We have (remember $\nu>0$)
$$\Psi_+(\nu;c, \alpha)~=~ e^{i \alpha \nu /2}~ \psi_{\nu}(c)
\eqno\hbox{(4.29a)}$$
for $E<0,~a<0$, and
$$\Psi_+(\nu;c, \alpha)~=~ e^{-i \alpha \nu /2}~ \chi_{\nu}(c)
\eqno\hbox{(4.29b)}$$
for $E>0,~a>0$. On the other hand the solutions corresponding to
$l<0$ are
$$\Psi_-(\nu;c, \alpha)~=~ e^{-i \alpha \nu /2}~ \psi_{\nu}(c)
\eqno\hbox{(4.30a)}$$
for $E>0,~a<0$, and
$$\Psi_-(\nu;c, \alpha)~=~ e^{i \alpha \nu /2}~ \chi_{\nu}(c)
\eqno\hbox{(4.30b)}$$
for $E<0,~a>0$.
Solutions (4.29-30) are the gauge--fixed wave functions correspondent of
(4.8). Analogously to the KG case the use of both positive and negative
$l$ is appropriate if one reinterprets the wave function as a quantum
operator (second quantization of BH). For instance,
$$\Psi_{\rm BH}(\alpha,c)=\int_0^\infty d\nu~\sqrt{{2 \nu
\sinh\pi\nu\over\pi
^2}}~K_{i\nu}(c)~[A^\dagger(\nu) e^{-i\nu\alpha/2} + B(\nu)
e^{i\nu\alpha/2}]\eqno(4.31)$$
is the representation of the BH quantum field for $a<0$.
\beginsection 5. Conclusions.
The quantization of the canonical approach to the black hole
proposed in I
shows that, as a consequence of the positive definiteness of the
canonical variable $b$, $\hat J$  does not have  a self--adjoint
extension since its conjugate variable $I$ has positive support.
Instead, eigenfunctions of $\hat J ^2$ can be defined in the Hilbert
space.

This possibly signals that the identification of $J$ with the mass
carried at the classical level is not the correct one in the quantum
formulation. Alternatively, this may have something to do with the fact
that in classical physics only positive masses are present. To look into
this question in the present frame one has to construct a procedure of
classical limit that yields the Schwarzschild metric and investigate the
role of eigenfunctions of $\hat J^2$. Maybe some light could come.

Another subject that must be explored is the introduction of matter
fields. This could be of importance in order to specify the physical
degrees of freedom inaccessible for observation by an external observer,
whose tracing out could explain the origin of the black hole entropy
(see e.g.\ [13]).
Hopefully, this may also shed light on the quantum definition of the mass
of the black hole.

The set of solutions of wormholes for the
KS metric  coincides with the set of Schwarzschild wave
functions inside the black hole, as the KS geometry
coincides with the internal one of the black hole, and the parameter
$r$ in which we foliate is timelike there.

Let us also remark that no quantization of the mass
appears from this theory. It is interesting to stress
though that in the frame developed here quantization of the mass squared
could be
achieved in a gauge invariant way by a  modification of the theory. For
instance a very crude way is just to set the support condition $x<x_0$.
This is a gauge invariant cut--off that leads to quantization of the
eigenvalues of $\hat J ^2$. Now, this cut--off is performed in the
gauge $y=1$, that is $2bx=1$. Thus a modification of the theory for large
$x$ corresponds to a gauge invariant modification for small $b$. It will
be interesting to explore the consequences of less crude models leading
to quantization of the mass; this requires a reliable definition of
the quantum mass operator of course.
\beginack
We are indebted to L.J. Garay for interesting discussions and useful
suggestions. One of the authors (A.T.F.) acknowledges a partial support
for this investigation from RFFI (grant 95-12-2a/154), ISF (grant RFF
300), and INFN.
\beginappendix A.
In this appendix we discuss the orthonormality of the eigenfunctions of
$\h N$ that we used in section 4 (Eqs.\ (4.27)). Let us start considering
$a<0$.
Using
$$\int_0^\infty {dx\over x} K_{i\mu}(x)~K_{i\nu}(x)=
{\pi^2\over 2\mu\sinh\pi\mu}\left[\delta(\mu-\nu)+\delta(\mu+\nu)\right]\,,
\eqno\hbox{(A.1)}$$
(see [14]) and recalling that $\nu$ is positive, we obtain (4.27a).

Let us discuss now in detail the case $a>0$. The most general solution
of the Eq.\ (4.25) has the form
$$\chi_\nu(\alpha,c)=\lambda_1 H_{i\nu}^{(1)}(c)+
\lambda_2 H_{i\nu}^{(2)}(c)\,,\eqno\hbox{(A.2)}$$
where $\lambda_{1,2}$ have to be determined by orthonormality.
We have to compute the integrals:
$$I^{(k,l)}(\mu,\nu)=\int_0^\infty {dx\over x} H_{i\mu}^{(k)}(x)~
H_{i\nu}^{(l)}(x)\,,\eqno\hbox{(A.3)}$$
where $k,l=1,2$. From Bateman (see Ref.\ [11], Vol.\ I, p.\ 333, formulae
(40) and (48)) we have the relation:
$$\int_0^\infty {dx\over x} H_{\mu}^{(k)}(x)~
H_{\nu}^{(k)}(x)=-{4\over\pi^2}e^{i\pi(2k-3)(\mu+\nu)/2}
\int_0^\infty {dx\over x} K_{\mu}(x)~K_{\nu}(x)
\,,\eqno\hbox{(A.4)}$$
where $k=1,2$. Setting $\mu\to i\mu$, $\nu\to i\nu$ in (A.4) and
using (A.1) we obtain
$$\eqalignno{&I^{(1,1)}(\mu,\nu)=-{2\over \mu\sinh\pi\mu}e^{\pi(\mu+\nu)/2}
\left[\delta(\mu-\nu)+\delta(\mu+\nu)\right]\,,&\hbox{(A.5a)}\cr
&I^{(2,2)}(\mu,\nu)=-{2\over \mu\sinh\pi\mu}e^{-\pi(\mu+\nu)/2}
\left[\delta(\mu-\nu)+\delta(\mu+\nu)\right]\,.&\hbox{(A.5b)}\cr}$$
Now, let us calculate $I^{(1,2)}$. In order to do this we
have to compute $\int_0^\infty J_{i\mu} J_{i\nu} dx/x$ and
$\int_0^\infty J_{i\mu} Y_{i\nu} dx/x$. These integrals can be
easily calculated using Bateman (see Ref.\ [11], Vol.\ I, p.\ 331-332,
formulae (33) and (36)) and suitable analytic continuations. We have:
$$\eqalignno{&\int_0^\infty {dx\over x}J_{i\mu}
J_{i\nu}=-{2i\over\pi}P{\sinh{[\pi(\mu-\nu)/2]}\over
\mu^2-\nu^2}+{1\over\mu}\sinh{\pi\mu}~\delta(\mu+\nu)\,,&\hbox{(A.6a)}\cr\cr
&\int_0^\infty {dx\over x}J_{i\mu} Y_{i\nu}=
{2\over\pi}P{\cosh{[\pi(\mu-\nu)/2]}\over \mu^2-\nu^2}+
{i\over\mu}\;\delta(\mu-\nu)+&\cr
&\hbox to 71 truemm{}+{i\over\mu}\cosh{\pi\mu}~\delta(\mu+\nu)
\,.&\hbox{(A.6b)}\cr}$$
Hence, using (A.5-6) we find:
$$I^{(1,2)}(\mu,\nu)=-{4i\over\pi}P{e^{\pi(\mu-\nu)/2}\over
\mu^2-\nu^2}+{2\over\mu}\,\hbox{coth}\,\pi\mu~\left[\delta(\mu-\nu)+e^{\pi\mu}
\delta(\mu+\nu)\right]\,.\eqno\hbox{(A.7)}$$
Now, we can calculate $\lambda_{1,2}$ imposing the inner product (4.27b).
We have two sets of real orthonormal functions:
$$\eqalignno{&\chi^{(1)}_{\nu}=\sqrt{\nu\cosh(\pi\nu/2)\over 4\sinh(\pi\nu/2)}
\left[e^{-\pi\nu/2}H^{(1)}_{i\nu}(c)+e^{\pi\nu/2}H^{(2)}_{i\nu}(c)\right]\,,
&\hbox{(A.8a)}\cr\cr
&\chi^{(2)}_{\nu}=i\sqrt{\nu\sinh(\pi\nu/2)\over 4\cosh(\pi\nu/2)}
\left[e^{-\pi\nu/2}H^{(1)}_{i\nu}(c)-e^{\pi\nu/2}H^{(2)}_{i\nu}(c)\right]\,.
&\hbox{(A.8b)}\cr}$$
In (4.26) we have chosen the set (A.8b) because it has the same
properties as (4.24), i.e.\ the wave functions vanish for $\nu\to 0$.
Also, the asymptotic behaviors for $c\to 0$ of (4.24) and (A.8b) are
identical.
\beginappendix B.
In this Appendix we collect the main formulae of sections 2--4 when the
cosmological constant $\Lambda$ is different from zero.

The system described by the Hamiltonian (2.7b) with $\Lambda\not=0$
(hereafter denoted $\tilde H$ to distinguish it from the Hamiltonian
$H$ of the previous sections) is again completely integrable. The finite
gauge transformations are in this case:
$$\eqalign{&b \ra \bar b=b+h(r)~{{\pa}\over{2b}}\,,\cr
&\pa \ra \bar \pa = \pa + h(r)~ {{\pa^2}\over{2b^2}}\,,\cr
&a \ra \bar a = a + {{\tilde N}\over{b^2}}~ {{h(r)/2}\over{1+h(r)\pa/2b^2}}
{}~+~{\Lambda\over 3}{b^2\over p_a}~h(r)
\left(1+{h(r)p_a\over 4b^2}\right)\,,\cr
&p_b \ra \bar p_b = p_b + {\tilde J\over b^2}~ {{h(r)/2}\over{1+h(r)\pa/2b^2}}
{}~+~{4\Lambda\over 3}bh(r)\left(1+{h(r)p_a\over 4b^2}\right)\,,\cr
&l(r) \ra \bar l(r) = l(r) + {{dh}\over{dr}}\,,\cr}\eqno\hbox{(B.1)}$$
where we have used the constraint $\tilde H=0$. The gauge invariant
quantities $\tilde J$ and $\tilde N$ are defined as:
$$\eqalignno{&\tilde J~=~J-{2\Lambda\over 3}b^3~
=~2b-p_ap_b+4b\tilde H+{4\over 3}\Lambda b^3\,,&\hbox{(B.2a)}\cr
&\tilde N~=~N-{2\Lambda\over 3}{b^4\over p_a}~=~bp_b - 2ap_a-
{2\Lambda\over 3}{b^4\over p_a}\,.&\hbox{(B.2b)}\cr}$$
The quantities (B.2a,b), together with
$$\tilde I~=~\tilde N\tilde J^{-1}~=~{b\over\pa}\,,\eqno\hbox{(B.2c)}$$
satisfy the algebra (2.12).
Note that the gauge transformations for $b$ and $p_a$ are unaffected by
the presence of the cosmological constant.

The rigid transformations generated by $\tilde I$
are identical to the ones generated by $I$ when $\Lambda=0$;
the finite transformations generated by the dilatation
operator $\tilde N$ are instead:
$$\eqalign{&b \ra \bar b = e^g b\,,\cr
&\pa \ra \bar \pa = e^{2g}\pa\,,\cr
&a \ra \bar a = e^{-2g} a+{\Lambda\over 3}{b^4\over
\pa^2}\left(1-e^{-2g}\right)
\,,\cr
&p_b \ra \bar p_b = e^{-g} p_b+{8\Lambda\over
3}{b^3\over\pa}\left(1-e^{-g}\right)
\,.\cr}\eqno\hbox{(B.3)}$$
Note that the presence of the cosmological constant does not affect the
rigid transformations of $b$ and $p_a$, again as it happens for the gauge
transformations (B.1). As a consequence, the discussion about the gauge
and the rigid invariant measure of the section 3 is applicable, as well as
the FP gauge fixing method described there.

For sake of completeness, let us write the gauge invariant relation
between $a$ and $b$:
$$a=\tilde I^2 \biggl(2\tilde H+1-{\tilde J\over b}+
{\Lambda\over 3}b^2\biggr)\,.\eqno\hbox{(B.4)}$$
Of course, $\tilde I$ and $\tilde J$ have the same physical meaning of
$I$ and $J$ in (2.13c).

In the $\{b,p_a\}$ representation the solutions of the WDW equation,
eigenfunctions of $\tilde N$, are:
$$\Psi_{\nu}(b,p_a)~=~c(\nu)~\left({\pa\over b}\right)^{i\nu}~
\exp\left[i{b^2\over\pa}\left(1+{\Lambda\over 3} b^2\right)\right]
\,.\eqno\hbox{(B.5)}$$
The eigenfunctions of the mass operator $\tilde J$ are instead:
$$\Psi_j(b,\pa)~=~c(j)\sqrt{{b\over\pa}}~
\exp\left\{i{b\over\pa}\left[b\left(1+{\Lambda\over
3}b^2\right)-j\right]\right\}\,.\eqno\hbox{(B.6)}$$
In the $\{\tilde J,\tilde I,\tilde Y\equiv Y,\tilde P_Y\equiv\tilde H\}$
representation (B.5) reads:
$$\Psi_{\nu}(\tilde x,\tilde y)~=~c(\nu)~\tilde x^{-i\nu}
\exp\left[i{\tilde y\over 2}\left(1+{\Lambda\over 12}
{\tilde y^2\over\tilde x^2}\right)\right]\,.\eqno\hbox{(B.7)}$$
Fixing the gauge $\Phi(\tilde y)=\tilde y-1=0$, and using the measure (3.2)
(where of course $x\to\tilde x$, $y\to\tilde y$), the orthonormal gauge fixed
wave functions are then:
$$\psi_{\nu}(\tilde x)~=~{1\over\sqrt{2\pi}}\tilde x^{-i\nu}
\exp\left(i{\Lambda\over 24\tilde x^2}\right)\,.\eqno\hbox{(B.8)}$$
To conclude this Appendix, let us find the eigenfunctions of $\tilde N$
and $\tilde J$ in the $\{a,b\}$ representation, using the Fourier
transform
(4.10). For the eigenfunctions of $\tilde N$, from (B.5) one obtains:
$$\Psi_\nu(a,b)=c(\nu)\left[{|a|\over 1+\Lambda
b^2/3}\right]^{-i\nu/2}B_{i\nu}\left(2b\sqrt{|a|(1+\Lambda
b^2/3)}\right)\,,\eqno\hbox{(B.9)}$$
where $B_{i\nu}=K_{i\nu}$ for $a<0$ and $B_{i\nu}=C_{i\nu}$ for $a>0$.
Analogously, from (B.6) the eigenfunctions of $\tilde J$ are
$$\Psi_j(a,b)=
{K(j)\over\sqrt{|b(1+\Lambda b^2/3)-j|}}\exp\left[\pm 2i\sqrt{ab[b(1+\Lambda
b^2/3)-j]}\right]\eqno\hbox{(B.10a)}$$
in the classically allowed region, and
$$\Psi_j(a,b)= {K(j)\over\sqrt{|b(1+\Lambda b^2/3)-j|}}\exp\left[-2\sqrt{ab[j-
b(1+\Lambda b^2/3)])}\right]\eqno\hbox{(B.10b)}$$
in the classically forbidden region. It is straightforward to verify
that Eqs.\ (B.9,10) satisfy the WDW equation and are respectively
eigenfunctions of $\tilde N$ and $\tilde J$ in the $\{a,b\}$ representation.
\beginref
\ref [1] H.A. Kastrup and T. Thiemann, \NPB{425}{1994}{665}.

\ref [2] K.V. Kucha\v r, \PRD{50}{1994}{3961}.

\ref [3] M. Cavagli\`a, V. de Alfaro and A.T. Filippov,
Hamiltonian Formalism for Black Holes and Quantization, to appear in:
Int.\ J.\ Mod.\ Phys.\ D (1995) and references therein.

\ref[4] M.P. Jr. Ryan, Hamiltonian Cosmology (Springer--Verlag, Berlin, 1972).

\ref [5] M. Cavagli\`a, \MPLA{9}{1994}{1897}.

\ref [6] S.W Hawking, \PRL{69}{1992}{406} and references therein.

\ref [7] See for instance: M. Henneaux and C. Teitelboim,
Quantization of Gauge Systems (Princeton University Press, Princeton,
New Jersey, 1992).

\ref [8] M. Cavagli\`a, V. de Alfaro and A.T. Filippov,
\IJMPA{10}{1995}{611}.

\ref [9] See for instance: A. Messiah, M\'ecanique Quantique
(Dunod, Paris, 1959).

\ref [10] Bateman Manuscript Project, Higher Transcendental
Functions, ed.\ A. Er\-d\'e\-lyi, Vol.\ II (McGraw--Hill Book Company Inc.,
New York, 1953).

\ref [11] Bateman Manuscript Project, Tables of Integral
Transforms, ed.\ A. Erd\'elyi (McGraw--Hill Book Company Inc., New
York, 1954).

\ref [12] J.D. Bjorken and S.D. Drell, Relativistic Quantum
Mechanics (Mc Graw--Hill Book Company Inc., New York, 1964).

\ref [13] A.O. Barvinsky, V.P. Frolov and A.I. Zelnikov, \PRD{51}{1995}{1741}
and references therein.

\ref [14] U. Gerlach, \PRD{38}{1988}{514}.

\endref
\bye